\documentstyle[epsfig]{mn}

\def\Msun{\ifmmode{~{\rm M}_\odot}\else${\rm M}_\odot$\fi}
\def\Lsun{\ifmmode{~{\rm L}_\odot}\else${\rm L}_\odot$\fi}
\def\kms{\ifmmode{$~km\thinspace s$^{-1}}\else km\thinspace s$^{-1}$\fi}

\def\etal{et al~}
\def\ee{\end{equation}}
\def\be{\begin{equation}}

\title{On the mass-to-light ratio of the local Galactic disc and the optical
luminosity of the Galaxy}

\author[Chris Flynn, Johan Holmberg, Laura Portinari, Burkhard Fuchs and 
Hartmut Jahrei\ss] 
{Chris Flynn$^{1,3}$, Johan Holmberg$^{1,4}$,  Laura Portinari$^{1}$,
Burkhard Fuchs$^{2}$, Hartmut Jahrei\ss$^2$\\ 
$^1$Tuorla Observatory, V\"ais\"al\"antie 20, FI-21500 Piikki\"o, Finland\\
$^2$Astronomisches Rechen-Institut am Zentrum f\"ur Astronomie der
Universit\"at Heidelberg,
M\"onchhofstrasse 12-14, Heidelberg, Germany\\
$^3$Mount Stromlo Observatory, Weston Creek, ACT, Australia\\
$^4$Max-Planck-Institut f\"ur Astronomie, K\"onigstuhl 17, Heidelberg, Germany}

\date{\today}

\begin{document}

\maketitle

\voffset=-1.0cm     

\begin{abstract} 
We measure the volume luminosity density and surface luminosity density
generated by the Galactic disc, using accurate data on the local luminosity
function and the disc's vertical structure. From the well measured volume mass
density and surface mass density, we derive local volume and surface
mass-to-light ratios for the Galactic disc, in the bands $B$, $V$ and $I$.  We
obtain mass-to-light ratios for the local column of stellar matter of $(M/L)_B
= 1.4 \pm 0.2$, $(M/L)_V = 1.5 \pm 0.2$ and $(M/L)_I = 1.2 \pm 0.2$. The
dominant contributors to the surface luminosity in these bands are main
sequence turn-off stars and giants. Our results on the colours and
mass-to-light ratios for the ``Solar cylinder'' well agree with population
synthesis predictions using Initial Mass Functions typical of the Solar
Neighbourhood. Finally we infer the global luminosity of the Milky Way, which
appears to be under-luminous by about 1-$\sigma$ with respect to the main locus
of the Tully--Fisher relation, as observed for external galaxies.
\end{abstract}


\section{Introduction}

The mass--to--light ($M/L$) ratio\footnote{Unless otherwise explicitly stated,
this paper always refers to the $M/L$ of stellar matter} is an important
constraint for studies of stellar populations and for chemo--photometric models
of the Milky Way; these often serve as a calibration point for the modeling of
disc galaxies in general (e.g.\ Boissier \& Prantzos 1999).  The surface
luminosity and mean colours of the Solar column are also useful comparison
points in extragalactic studies, and for placing the Milky Way on the
Tully-Fisher relation.

Studies of the Galactic disc over the last two decades have resulted in good
determinations of its local mass density $\rho(0)$, and the surface mass
density, $\Sigma_0$ (e.g.\ Holmberg \& Flynn 2000, 2004 and references
therein). In this paper we study the related issue of the luminosity generated
by the local disc, both in the local volume (i.e.\ the volume luminosity
density) and integrated perpendicularly to the disc in a column (i.e.\ the
surface luminosity density). These quantities allow us to measure the
mass-to-light ratio, $(M/L)$, for the local Galactic disc.

Estimating the luminosity surface density (i.e.\ surface brightness) of the
local Galactic disc requires good knowledge of its vertical structure.  In this
respect, Galactic models have much improved since the 1980s, as a particular
result of the Hipparcos satellite and star count programs made with the Hubble
Space Telescope.  These data allow us to overcome the main deficiency in
earlier studies of the local disc surface brightness and disc $M/L$; due to the
lack of accurate distances to individual stars, the column luminosity (and
mass) density had to be recovered via assumptions about the scale-length $h_R$
and scale-height $h_z$ of the stellar disc, and the results were degenerate 
with respect to the assumed $h_z/h_R$ ratio. Distances to stars from Hipparcos 
now allow us to determine directly the distribution of stellar scale-heights 
and measure column densities, merely by ``counting'' stars and light at the
Galactic poles; $h_R$ is no longer required in the modeling and the degeneracy
is broken.

The infrared structure of the Milky Way has been extensively studied in the
1990s, mostly taking advantage of the COBE/DIRBE experiment, and is presently
well understood (Kent et~al.\ 1991; Dwek et~al.\ 1995; Binney et~al.\ 1997;
Freudenreich 1998; Bissantz \& Gerhard 2002). On the other hand, 
in the optical,
literature estimates of the disc's luminosity and $M/L$ ratio can mostly be
traced back to work done prior to the launch of both the Hipparcos satellite
and Space Telescope (see Table~\ref{tab:literature} for a list of the main
references). The time is thus ripe for a redetermination of the disc's
mass-to-light ratio.

\begin{table*}
\caption{Previous determinations of the surface brightness and colours of
the local Galactic disc}
\begin{tabular}{c c c c c|c c c| p{4truecm}}
\hline
\multicolumn{2}{c}{Solar cylinder} & \multicolumn{3}{c}{Disc} &
\multicolumn{3}{c}{Galaxy} \\
$\mu_B (R_{\odot})$ & $\mu_V (R_{\odot})$ & $M_{B,d}$ & $M_{V,d}$ & 
$(B-V)_{0,d}$ & $M_{B,g}$ & $M_{V,g}$ & $(B-V)_{0,g}$ & references \\
\hline
24.15$\pm$0.07 & & --19.61$\pm$0.06 & & 0.40 & 
--20.08$\pm$0.04 & & 0.53 $\pm$0.02 & de Vaucouleurs \& Pence (1978)\\
 & $\sim$23 & --19.9 & --20.4 & 0.45 &
--20.5 & --20.1 & 0.45 & Bahcall \& Soneira (1980), \\
 & & & & & & & & Bahcall (1984) \\
23.3$\pm$0.3  & 22.7$\pm$0.2 & & & 0.62$\pm$0.04 & & & & 
Ishida \& Mikami (1982) \\
23.8$\pm$0.1 & & --20.2$\pm$0.2 & & 0.84$\pm$0.15 & 
--20.3$\pm$0.2 & & 0.83 $\pm$0.15 & van der Kruit (1986) \\
\hline
\end{tabular}
\label{tab:literature}
\end{table*}

In sections 2 and 3 we determine the disc luminosity and mass-to-light ratio,
both locally and in a column at the Sun, for the optical bands $B$, $V$ and
$I$.  In section~4, we compare our results to theoretical predictions from
population synthesis modeling. In section 5, we compute the total disc
luminosity, for a range of plausible scale-lengths, and make some comparisons
between the Milky Way and external galaxies.  In section 6 we summarise and
discuss our results. The present study focuses on optical bands, and work is
underway to extend these measurements to near infra-red bands.

\section{Two independent studies}

The luminosity of the local Galactic disc has been determined in this paper, by
starting with the stellar luminosity function, and computing the total
luminosity the stars contribute both locally, and in a column integrated above
and below the Sun's position in the disc. From the total luminosity, and the
known disc mass density and surface density, we then derive disc mass-to-light
ratios.

Our analysis of the luminosity budget for the disc has been carried out
independently by the group working at Tuorla (CF, JH, LP) and the other at
Heidelberg (BF, HJ); we shall hereafter refer to the two programs as the Tuorla
and Heidelberg studies. We found out about each other's studies when they were
essentially completed, at which point we decided to combine efforts and discuss
the results together. As will be seen, the two studies were in excellent
agreement.

In the Tuorla study, our approach was to leverage existing work, carried out
some years earlier, on the disc's mass density. In Holmberg, Flynn \& Lindegren
(1997) and Holmberg \& Flynn (2000, 2004), we constrained the vertical
structure of the disc over a wide range of stellar types, primarily using the
Hipparcos and Tycho surveys. It was relatively straightforward to use those
calibrated models of the local Galactic disc to derive its volume and surface
luminosity density.

The Heidelberg study is based on very extensive work on the local stellar
luminosity function obtained from the CNS4 (Catalogue of Nearby Stars).  The
high quality colour, luminosity and velocity data for the sample stars were
used to compute the disc's volume luminosity density and surface luminosity
density.

The essential difference between the two studies is that the Tuorla sample
reaches to more luminous stellar types, because it surveys a deeper volume
(that probed by Hipparcos/Tycho out to circa 200 pc); and involves a model of
the local disc structure. The Heidelberg study reaches to much lower luminosity
stars, and is limited to a distance of 25--50~pc from the Sun; furthermore, it
is based on completely empirical local stellar data and relies on no
modeling. The agreement between the Tuorla and Heidelberg studies, where they
overlap, turned out to be excellent, and we have combined the results of the
studies with confidence.

\subsection{The Tuorla study}

The disc luminosity calculations in the Tuorla study are based on a description
of the local disc (Holmberg \& Flynn 2000, 2004), composed of both gaseous and
stellar components, and constructed for the purpose of determining the disc's
vertical mass distribution.

\begin{figure}
\begin{center}
\includegraphics[width=0.35\textwidth,angle=-90]{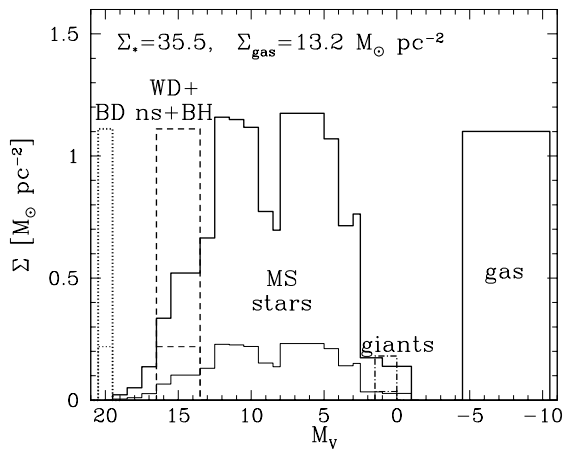}
\end{center}
\caption{The Disc Mass Model. For each component of the model, the total
contribution by mass is shown as a function of the $V$-band absolute
magnitude. MS stands for main sequence, WD+ns+BH for white dwarfs, neutron star
and black hole type stellar remnants (their luminosity is dominated by the
WDs), BD for brown dwarfs. ``Giants'' are first ascent and helium core burning
stars of about a solar mass. Thick lines show the total mass contribution
(thin+thick disc), thin lines show the thick disc fraction.  Note that the
gaseous components have been included on the right hand side of the figure
(gas), but do not generate optical luminosity). Full details of the model,
represented graphically here, are in Table \ref{table:massmodel}.  The mass
model actually assigns one single value to the global surface mass density in M
dwarfs ($M_V>8$), based on HST studies; for the sake of this plot, such global
mass in faint stars has been distributed in magnitude following the mass
function of Binney \& Merrifield (1998).}
\label{fig:massmodel}
\end{figure}

Fig.~\ref{fig:massmodel} shows the mass contributions made by the components of
the model for the disc. It is an updated version of the mass model of Holmberg
\& Flynn (2000), and is shown in Table \ref{table:massmodel}. Full details of
how these models are constructed can be found in Holmberg \& Flynn (2000,
2004). The models consists of a thin disc and a thick disc (and a stellar halo
as well, although this is irrelevant in the present study).

The stellar components of the model consist, broadly speaking, of main sequence
stars of different $M_{\rm V}$ (indicated by MS in Fig.~\ref{fig:massmodel}),
red giants (i.e. first ascent and He core burning giants of about a solar
mass), supergiants (i.e. relatively massive, luminous giants), white
dwarfs/neutron stars/black holes (i.e. stellar remnants, indicated by WD+ns+BH)
and brown dwarfs (indicated by BD in the figure). The scale-heights (i.e. the
density falloff with vertical height above the disc) of each stellar component
have been constrained by star-count data from the Hipparcos and Tycho catalogs
(for stars brighter than M dwarfs), or via Space Telescope (for the M
dwarfs). The scale-heights for stars dimmer than the main sequence turn-off are
mainly constrained by self-consistency with the mass model (via their known
velocity dispersions and the Poisson-Boltzmann equation).

In this model, the surface mass density in stars is 
{\mbox{$\Sigma_* = 35.5$ \Msun pc$^{-2}$} and for the gaseous components, 
the surface mass density is {\mbox{$\Sigma_{\mathrm gas} 
= 13.2$ \Msun pc$^{-2}$.}}

\begin{table}
\small
\caption{The Disc Mass Model. Mass components in the disc consist broadly of
gas, main sequence stars and giants, stellar remnants and substellar objects.
For each component, the table gives the local mass density at the Galactic
mid-plane $\rho(0)$, the vertical velocity dispersion $\sigma_W$, and the
surface mass density, $\Sigma$.}
\begin{center}
\begin{tabular}{llrr}
\hline
Description&$\rho(0)$~~~~~& $\sigma_W$~~~~~  & $\Sigma$~~~~~   \\
           & \Msun pc$^{-3}$ & km s$^{-1}$ &  \Msun pc$^{-2}$ \\
\hline
H$_2$               & 0.021   ~~~~&   4.0 ~~~~ &   3.0~~~~~~   \\
H\thinspace I(1)    & 0.016   ~~~~&   7.0 ~~~~ &   4.1~~~~~~   \\
H\thinspace I(2)    & 0.012   ~~~~&   9.0 ~~~~ &   4.1~~~~~~   \\
warm gas            & 0.0009  ~~~~&  40.0 ~~~~ &   2.0~~~~~~   \\
giants              & 0.0006  ~~~~&  20.0 ~~~~ &   0.4~~~~~~   \\
$M_V < 2.5$         & 0.0031  ~~~~&   7.5 ~~~~ &   0.9~~~~~~   \\
$2.5 < M_V < 3.0$   & 0.0015  ~~~~&  10.5 ~~~~ &   0.6~~~~~~   \\
$3.0 < M_V < 4.0$   & 0.0020  ~~~~&  14.0 ~~~~ &   1.1~~~~~~   \\
$4.0 < M_V < 5.0$   & 0.0022  ~~~~&  18.0 ~~~~ &   1.7~~~~~~   \\
$5.0 < M_V < 8.0$   & 0.007   ~~~~&  18.5 ~~~~ &   5.7~~~~~~   \\
$M_V > 8.0$         & 0.0135  ~~~~&  18.5 ~~~~ &  10.9~~~~~~   \\
white dwarfs        & 0.006   ~~~~&  20.0 ~~~~ &   5.4~~~~~~   \\
brown dwarfs        & 0.002   ~~~~&  20.0 ~~~~ &   1.8~~~~~~   \\
thick disk          & 0.0035  ~~~~&  37.0~~~~~ &   7.0~~~~~~   \\
stellar halo        & 0.0001  ~~~~& 100.0~~~~~ &   0.6~~~~~~   \\
\hline 
\end{tabular}
\end{center}
\label{table:massmodel}
\end{table}

For the stellar components in the model, we have computed how much light is
contributed to the local volume and local column. In the $V$ band, this is
straightforward, as all the stellar components in the model have well measured
absolute $V$-band luminosity, $M_V$, either from space-based parallax data
(i.e. Hipparcos, for $M_V < 8$) or ground-based parallax data (for $M_V > 8$).

To convert the $V$ luminosities to other bands ($B$ and $I$), we used the
$(B-V)$ and $(V-I)$ colour distributions in the Hipparcos and Tycho catalogues.
Only directly measured, ``a'' flagged $(V-I_c)$ colours were considered, for
homogeneity and accuracy. The transformations are very similar to what can be
obtained from the collated $UBVRI$ data for nearby stars with very accurate
parallaxes (Reid 2005, NSTAR catalogue, private communication).

To translate magnitudes to luminosities we adopt a $V$ band absolute magnitude
for the Sun of $M_{V,\odot} = 4.82$; for the other bands we adopt the solar
colours from Holmberg, Flynn \& Portinari (2006), $(B-V)_\odot=0.64$ and
$(V-I)_\odot=0.69$.

\subsection{The Heidelberg study}

At Heidelberg, the disc luminosity density was computed in a slightly different
manner. The starting point was the local disc luminosity function, obtained
from the CNS4 (Catalogue of Nearby Stars). 
This represents a census of stars within 25~pc (Jahrei\ss\ \& Wielen 
1997), extending to a 50~pc volume for the brightest stars 
(Jahrei\ss, Wielen \& Fuchs 1998; Table~\ref{tab:CNS4_50pc}). 
The CNS4 is complete within 25 pc for stars of spectral type K and earlier, 
but for later spectral types only in smaller counting volumes. The luminosity
function constructed from the star counts has been carefully corrected for
this incompleteness (Jahreiss \& Wielen 1997).

\begin{table}
\caption{Luminosity function from the 50~pc sample of the CNS4 catalogue,
improving and extending the luminosity function of Jahrei\ss\ \& Wielen 
(1997) in the brightest luminosity bins. $\Phi$ is expressed 
in terms of number of stars within a 20~pc volume; $\epsilon_\Phi$ is the
Poisson error; $\Phi$(MS) refers to main sequence stars only.}
\begin{tabular}{rrrrrr}
\hline
$M_V$ & N$_{50}$ & N$_{50}$(MS) & $\Phi$ & $\epsilon_\Phi$ & $\Phi$(MS) \\
\hline
 --3 &   1 &   1 & 0.06 & 0.06 & 0.06 \\
 --2 & --- & --- & ---  & ---  & ---  \\
 --1 &  17 &   9 &  1.1 & 0.3  & 0.6  \\
   0 &  51 &  26 &  3.3 & 1.5  & 1.7  \\
   1 & 176 &  96 & 11.3 & 0.9  & 6.1  \\
   2 & 261 & 224 & 16.7 & 1.0  & 14.3 \\
   3 & 552 & 483 & 35.3 & 1.5  & 30.9 \\
\hline
\end{tabular}
\label{tab:CNS4_50pc}
\end{table}

\begin{figure}
\begin{center}
\includegraphics[width=0.49\textwidth]{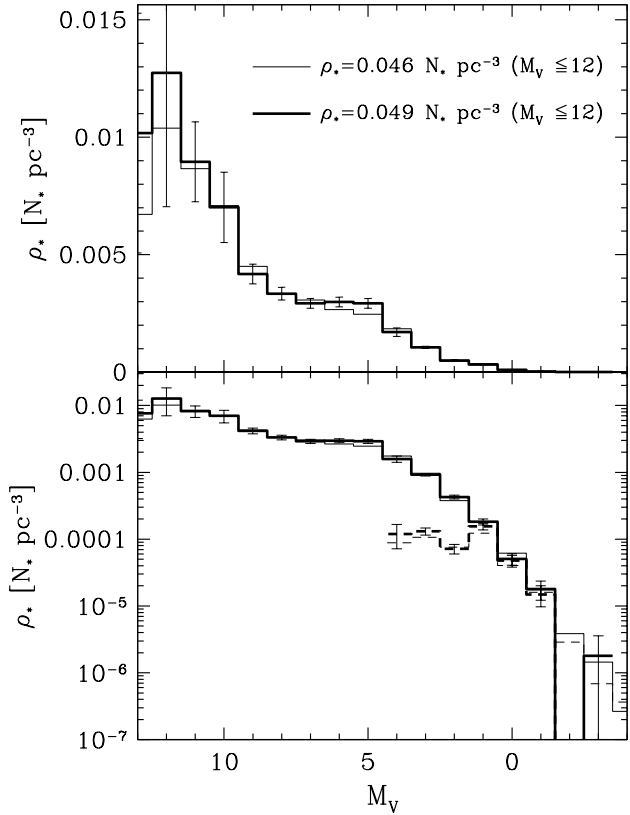}
\end{center}
\caption{Luminosity function of stars near the Sun, in units of pc$^{-3}$
mag$^{-1}$. In this and following figures, the thin curve shows the Tuorla
study, and the thick curve with error-bars the Heidelberg study, the two having
been carried out independently. The Tuorla study is based on star counts and
fits to the mass distribution of the local Galactic disc due to Holmberg \&
Flynn (2000), and probes out to 200 pc from the Sun.  The Heidelberg study is
based on the CNS4 (Catalogue of Nearby Stars), and is limited to stars within
25--50 pc from the Sun. Poisson error bars are shown for the Heidelberg data;
the Tuorla error bars are as small or smaller than the Heidelberg ones and are
not shown for clarity.  {\it Upper panel}: total luminosity function; {\it
lower panel}: luminosity function for main sequence stars (solid lines) and red
giants (dashed lines) separately.}
\label{fig:lfN}
\end{figure}

Estimating the local luminosity density from the LF is straightforward.  For
the surface luminosity density we proceeded as follows.  For stars in a given
magnitude bin, an observational vertical velocity dispersion $\sigma_W$ can be
obtained directly from the known space velocities (Jahrei\ss\ \& Wielen 1997).
The detection probability of a star in the volume is $P \propto \sigma_W^{-1}$
henceforth each star type is assigned a weight $1/P \propto \sigma_W$ when
going from volume to column quantities (cf.\ Fuchs \etal 2001), i.e.\ for each
magnitude bin:

\begin{equation} 
\label{eq:Sigma_fuchs}
\Sigma_L (M_V) \propto \rho_L (M_V) \times \sigma_W(M_V)
\end{equation}

\begin{figure}
\begin{center}
\includegraphics[width=0.35\textwidth]{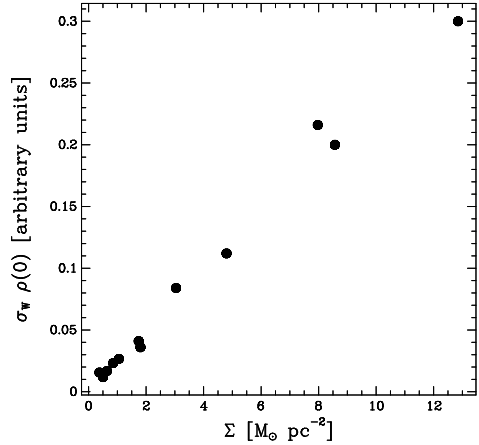}
\end{center}
\caption{Comparison of the surface density of individual components in our disc
models, versus the quantity $\sigma_W \times \rho(0)$ (i.e. the product of
vertical velocity dispersion and the density of the component at the disc
mid-plane). This shows that Eqn \ref{eq:Sigma_fuchs} holds to a a very good
approximation for realistic discs.}
\label{fig:rhosig}
\end{figure}

We have verified that this relationship holds in realistic discs, by computing
$\rho_* (M_V) \times \sigma_W(M_V)$ for each component of the Tuorla disc
model, and comparing it to the surface density found for each component after
solving for the density falloff of all components fully self-consistently via
the Poisson-Boltzmann equations. As can be seen in Fig.~\ref{fig:rhosig} the
quantities are very closely proportional. Eq.~\ref{eq:Sigma_fuchs} is used in
the next section to compare the Tuorla and Heidelberg results on surface
luminosity density.

Fig.~\ref{fig:lfN} shows the luminosity function from CNS4 (thick lines with
error-bars) compared to the one used in the Tuorla Galactic model (thin lines);
the agreement is excellent, for both main sequence and giant stars (lower
panel, dashed lines). We now compute the luminosity budget for the local
disc based on the Tuorla and Heidelberg studies.

\subsection{Local disc $V$-band luminosity budget}

We begin with the $V$-band, since this computation was most straightforward.

The local volume luminosity density, $\rho_L$ in the $V$-band is shown as a
function of $V$ absolute magnitude, $M_V$, in Fig.~\ref{fig:lf_V} for the two
studies (Tuorla: thin lines; Heidelberg: thick lines with error-bars).  The top
panel shows the total luminosity emitted by each magnitude wide bin from the
local stellar luminosity function; we further compare to the volume luminosity
density distribution of Binney \& Merrifield (1998; dotted line).  The Tuorla
study is able to constrain the contribution by luminous stars well because it
reaches deeper (about 200 pc) compared to the Heidelberg study (which is for
local stars out to 25--50 pc). This is why the Heidelberg data reliably probes
luminosities only for $M_V \geq -1$.  For less luminous stars, where we can
compare the two studies directly, the agreement is excellent. The luminosity
generated for stars with $M_V \geq -1$ is $0.045$ \Lsun pc$^{-3}$ in the Tuorla
study and $0.047$ \Lsun pc$^{-3}$ in the Heidelberg study.  From the Tuorla
study, we derive a total volume luminosity generated for all stars of $\rho_L =
0.056$ \Lsun pc$^{-3}$.

\begin{figure}
\begin{center}
\includegraphics[width=0.49\textwidth]{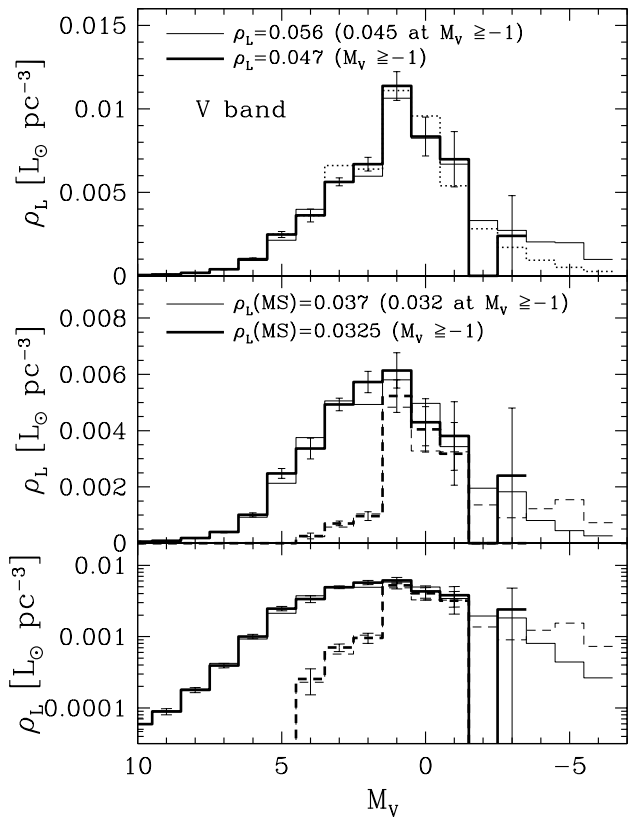}
\end{center}
\caption{Luminosity generated by stars in the $V$-band near the Sun, in units
of {\mbox{\Lsun pc$^{-3}$ mag$^{-1}$;}} as in Fig.~\protect{\ref{fig:lfN},
thin curves for the Tuorla study, thick curves with error-bars for the 
Heidelberg study.
{\it Top panel}: total $V$-band luminosity budget, peaking} in the local volume
in the range {\mbox{$-1 < M_V < 4$;}} also shown as a dotted line is the 
$V$-band luminosity distribution of Binney \& Merrifield (1998).  
{\it Middle and bottom panels}: luminosity budget separated into main sequence 
(solid lines) and giant components (dashed lines); in linear and logarithmic 
scale respectively.}
\label{fig:lf_V}
\end{figure}

One question we wished to address is which stars dominate the luminosity
budget, in the volume as well as in the column, and how is this a function of
wavelength (or colour band).  We examine this question by dividing the samples
into main sequence stars and giants and show the results in the middle panel of
Fig.~\ref{fig:lf_V}. The main sequence is represented by the solid lines, while
the giants are shown by the dashed lines. There is excellent agreement between
the samples; the bottom panel shows the same comparison but in logarithmic
scale, to highlight the excellent agreement down to the very faint main
sequence.

Fig.~\ref{fig:lf_V} shows clearly, and not surprisingly, that the main
contributors to the local $V$-band volume luminosity budget are the main
sequence stars around the turn-off, in the range {\mbox{$3 < M_V < 0$;}} giants
contribute mainly in the range {\mbox{$1 < M_V < 0$}} (location of the red 
clump).

The results so far have been for volume luminosity density, $\rho_L$. More
interesting, from the point of view of studies of external galaxies, is the
surface luminosity density $\Sigma_L$, to which we now turn.

In the Tuorla study, we sum the total contribution of thin and thick disc
components in the model by integrating in $z$, i.e. vertically in both
directions out of the disc, and accounting for the known falloff of the stars
as a function of height. In Fig.~\ref{fig:lfV_tuorla} we overplot the volume
and column luminosity densities for the Tuorla Galactic model.  The total
luminosity of the column, from the Tuorla data, is $\Sigma_L = 24.4$ \Lsun
pc$^{-2}$.

The comparison to the Heidelberg results is possible via the ``scaled'' surface
luminosity of Eq.~\ref{eq:Sigma_fuchs}, namely using the vertical velocity 
dispersion of the stars as a proxy for their scale-height in the potential;
we have shown in Fig.~\ref{fig:rhosig} that this is an excellent approximation
in the Tuorla Galactic model.

In the Heidelberg study, we adopt for main sequence stars the velocity
dispersions of Jahrei\ss\ \& Wielen (1997), and we further assign $\sigma_W =
23$ km/sec to old giants and 12 km/sec to clump giants (which are about half of
the giants in the magnitude bins $M_V=0-1$, Jahrei\ss, Fuchs \& Wielen
1999). Relatively low velocity dispersions for a significant fraction of red
giants were found also by Flynn \& Fuchs (1994); indeed synthetic
Hertzsprung--Russel diagram analysis shows that the majority of clump giants in
the local volume is expected to be as young as 1--2~Gyr (Girardi \& Salaris
2001).

Fig.~\ref{fig:lfV_col} clearly shows very good agreement between the two
studies also in the surface brightness estimate: the predicted overall surface
luminosity from stars with $M_V \leq -1$ (the common magnitude range in the
two studies) agrees to better than 10\%, with the agreement as good as a few
percent for the MS star contribution, and within 15\% for the giant
contribution when considered separately.

\begin{figure}
\begin{center}
\includegraphics[angle=-90,width=0.45\textwidth]{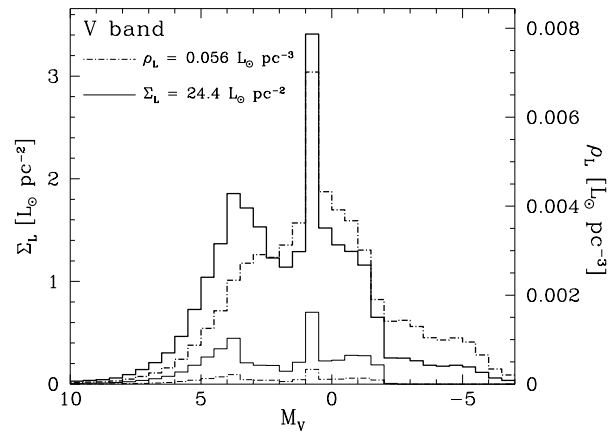}
\end{center}
\caption{V--band luminosity density from the Tuorla disc model in the solar
volume (dash--dotted line) and column (solid line); thick histograms for the
total (thin+thick) disc, thin lines for the thick disc fraction.  In the
surface luminosity density, two clear peaks are seen in the luminosity
contribution corresponding to turnoff main sequence stars and to clump giants.}
\label{fig:lfV_tuorla}
\end{figure}

\begin{figure}
\begin{center}
\includegraphics[width=0.45\textwidth]{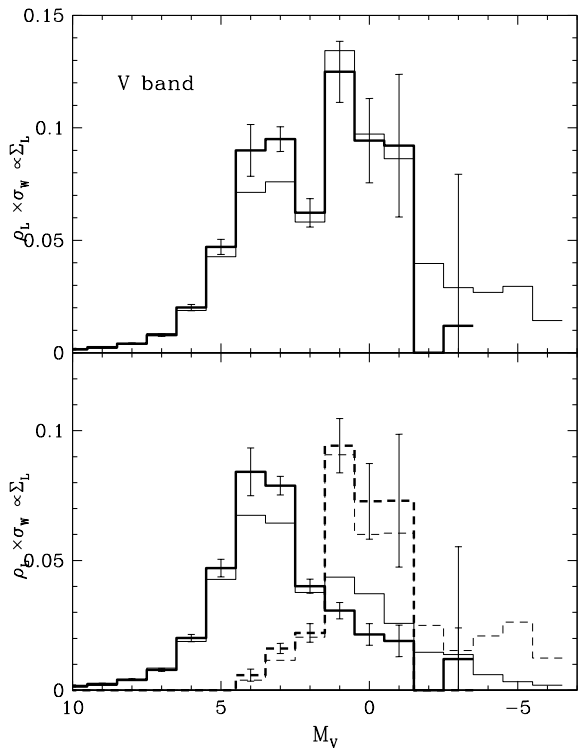}
\end{center}
\caption{Luminosity budget for the local disc in the $V$-band, shown by surface
density rather than volume density. We have used $\rho_L \times sigma_W$ as a
proxy for $\Sigma_L$, as discussed in the text. There is good agreement between
the two data sets. The Tuorla study results in a total $V$-band surface
luminosity of $\Sigma_L = 24.4$ \Lsun pc$^{-2}$.  {\it Bottom panel}: main
sequence (solid) and giant stars (dashed) separately. The two peaks in the
luminosity contribution, due to turnoff stars ($M_V \approx 3-4$) and giants
(mainly around $M_V \approx 0.5$), separate out clearly.  In the $V$-band,
these two components contribute about equally to the total light.}
\label{fig:lfV_col}
\end{figure}

Turning now to dominant contributors to the luminosity budget,
Figs.~\ref{fig:lfV_tuorla} and~\ref{fig:lfV_col} again show that the luminosity
comes mainly from stars in the range {\mbox{$4 < M_V < 0$}} --- 
i.e.\ from turn-off
stars, and from red giants near the ``clump'', as we shall see in more detail
below. The contribution from stars brighter than absolute magnitude $M_V = -1$,
i.e.\ very bright main sequence stars and giants, is only 2.3~\Lsun~pc$^{-2}$
--- or about 10\% of the total light in the column. This is a good deal less
than the 20\% quota they contribute to the volume luminosity (cf.\ solid
vs. dot--dashed line in Fig.~\ref{fig:lfV_tuorla}) their contribution being
suppressed in the column because bright young MS stars have low velocity
dispersion and low scale-height in the disc.

The bottom panel in Fig.~\ref{fig:lfV_col} shows the luminosity separated 
in main sequence and giant contributions. The dominant part of the luminosity, 
in the range {\mbox{$4 < M_V < 0$,}} is now seen to separate into two clean 
peaks --- 
one due to turnoff stars at $M_V \approx 3-4$ and the other due to giants 
(primarily ``clump'' giants, or core helium burning stars), at 
$M_V \approx 0.5$. In the $V$ band, the two contribute about equally 
(60--40\%, see Table~\ref{tab:details}) to the surface luminosity density, 
$\Sigma_L$.
 
The comparison between the Tuorla and Heidelberg results also highlights the
importance of the luminosity bins at $M_V = $3--4: the volume to column
transformation gives an important weight to these bins, and the differences in
this range between the Tuorla and the Heidelberg results is mainly due to the
slightly different $\sigma_W$'s adopted in the two studies in this range.  This
is the magnitude range where scale-height is most rapidly changing with
luminosity, and was the part of the Tycho and Hipparcos data which had to be
fit most carefully (Holmberg \etal 1997). Due to the scale-height and vertical
velocity dispersion effect, when going from volume to column luminosity
density, the peak of the MS stars luminosity contribution indeed shifts from
$M_V \sim 2$ (Fig. \ref{fig:lf_V}, mid panel) to $M_V$=3--4
(Fig. \ref{fig:lfV_col}, bottom panel). Also in Fig.~\ref{fig:lfV_tuorla},
overplotting the volume and column luminosity densities for the Tuorla Galactic
model, the peak of the main sequence contribution at $M_V$=3--4 emerges in the
surface luminosity density.

We now turn to the $M/L$ ratio of the disc in the $V$-band. The Tuorla Galactic
model yields in the local volume a $V$-band luminosity density $\rho_L = 0.056$
\Lsun pc$^{-3}$; the stellar mass density is $\rho_* = 0.042$ \Msun pc$^{-3}$
--- this yields a disc mass-to-light ratio in the $V$-band for the local volume
of $(M/L)_V = 0.75$~\Msun /\Lsun.  The local stellar density is determined by
Hipparcos data to better than 10\%, and 10\% is also the typical uncertainty in
the luminosity, as estimated from the ``freedom'' in adjusting the parameters
of the Tuorla model vs.\ all available observational constraints, and also
based on the comparison with the independent Heidelberg results
(Fig.~\ref{fig:lf_V}). Adding in quadrature, we estimate the uncertainty in
$M/L$ as $\approx 15$\%.

For the column, the luminosity surface density we derive is $\Sigma_L =
24.4$~\Lsun~pc$^{-2}$; the column density of {\it stellar} matter is
$35.5$~\Msun~pc$^{-2}$; thus, the surface mass-to-light ratio at the Sun's
position in the disc, in the $V$ band, is $(M/L)_V = 1.5$~\Msun /\Lsun, again
with an uncertainty of the order of 15\%.

In external galaxy studies, the mass-to-light ratio for all the {\it visible}
matter, i.e. including the gaseous component, is often the relevant quantity.
Adopting a local gas surface density of 13.2~\Msun~pc$^{-2}$ (uncertain by
about 50\%, Holmberg \& Flynn 2000 and references therein;
Table~\ref{table:massmodel}), the total surface density of visible baryons
comes to $\Sigma_{bar}$=48.7$\pm$7.5~\Msun~pc$^{-2}$ and the corresponding
total mass--to--light ratio increases by a factor of 1.4, to
{\mbox{$(M_{bar}/L)_V =2$~\Msun /\Lsun}} with an uncertainty of 20\%. This
applies to the local disc only; whenever possible in external galaxies one
considers the stellar and gaseous components separately, as our $M/L$ estimates
can be rigorously applied to the stellar component only and the gas fraction is
not universal.

Table~\ref{tab:luminosities} and Table~\ref{tab:details} summarize these
results (for $BVI$ bands) on the volume and surface brightness and colours for
the local thin disc, thick disc and total disc; the percentage of light
contributed by main sequence versus red giant stars; and the stellar
mass-to-light ratio ($M_*/L$) of the Solar cylinder.

\section{$B$ and $I$ band analysis}

We now proceed to the other bands --- $B$ and $I$. For the Heidelberg data, the
V band luminosity contributed by each magnitude bin was transformed into $B$
and $I$ using the average colour--magnitude relations from the HR diagram of
the CNS4 catalogue.  At Tuorla, we transformed the $V$ band luminosity
function, to which the Hipparcos/Tycho and HST star counts had been fit, to
other bands via colour-colour relations for stars in the Hipparcos/Tycho
catalogue, as discussed in Section 2.1.

\subsection{$B$-band disc luminosity and mass-to-light}

The $B$ band results are shown in Fig.~\ref{fig:lfMS_B}. Note that all our
results in the non $V$-bands are plotted as a function of $M_V$, to assist
comparison with the $V$-band analysis. As before, the Tuorla results are shown
by the thin lines and the Heidelberg results the thick lines (with
error-bars). The total volume luminosity density in the $B$ band, as determined
from the Tuorla sample, is $\rho_L = 0.074$~\Lsun pc$^{-3}$, most of which
(0.06~\Lsun pc$^{-3}$) from main sequence stars.  Up to $M_V = -1$, the
Tuorla and Heidelberg results for MS stars can be compared; the luminosity
density up to this point in the Heidelberg sample is $\rho_L = 0.048$~\Lsun
pc$^{-3}$, in excellent agreement with the Tuorla estimate. The luminosity
contributed by giants in B band is less than 20\% (Table~\ref{tab:details}).

\begin{figure}
\begin{center}
\includegraphics[width=0.35\textwidth,angle=-90]{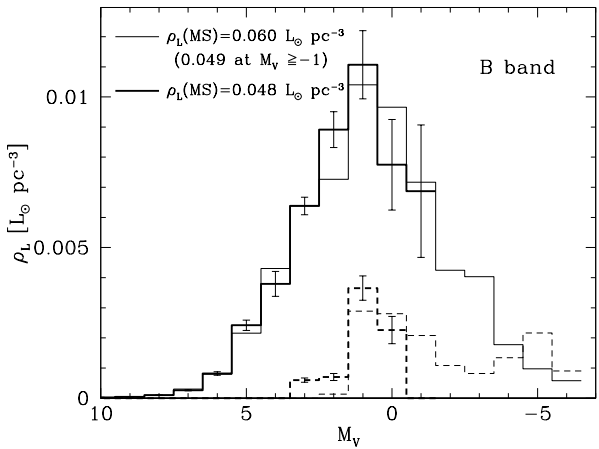}
\end{center}
\caption{Luminosity density in the local volume in the $B$-band. Tuorla results
as thin lines, Heidelberg results as thick lines with error-bars.  There is
excellent agreement between the samples for main sequence stars in the
overlapping magnitude range (solid lines); the dashed line is the luminosity
contribution from giants in the Tuorla model.}
\label{fig:lfMS_B}
\end{figure}

\begin{figure}
\begin{center}
\includegraphics[width=0.35\textwidth,angle=-90]{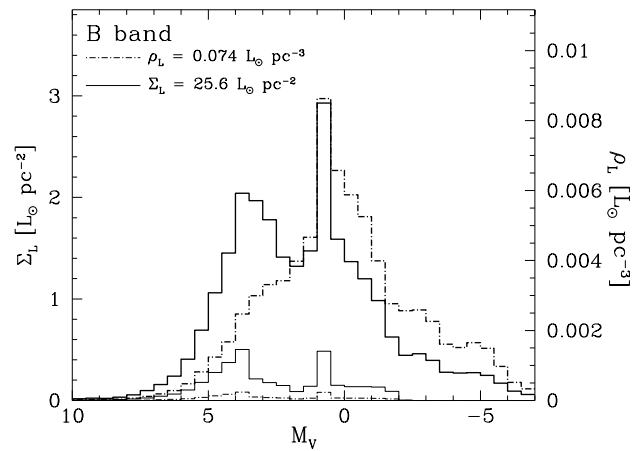}
\end{center}
\caption{Same as Fig.~\protect{\ref{fig:lfV_tuorla}}, in the $B$ band.  Turnoff
stars dominate the budget in the $B$-band, with the giants providing only 26\%
of the light, whereas the two components make about equal contributions to the
luminosity in the $V$ band.}
\label{fig:lfB_tuorla}
\end{figure}

The main luminosity contributor in the local volume in the $B$-band is from
stars at $M_V \approx 1$. Interesting results emerge when the surface
luminosity is computed, as seen in Fig.~\ref{fig:lfB_tuorla}.  We show only the
Tuorla data for clarity and completeness, as the Heidelberg sample does not
probe the luminosity function deeply enough.  The upper set of curves show both
disc components (thin+thick), while the lower set (thin lines) shows the thick
disc only.  The dash--dotted curves show the volume luminosity density
($\rho_L$); the solid curves show the surface luminosity density
($\Sigma_L$). As in the $V$ band, there are two clear peaks in the surface
luminosity contribution --- one at $M_V \approx 1$ due to clump giants and
bright main sequence stars and the other from turnoff stars ($M_V \approx
4$). Giants contribute only 26\% of the surface luminosity in $B$; in this
band, turnoff and bright main sequence stars provide most of the luminosity; in
the $V$ band instead the proportion is 40--60\%.

We obtain a volume luminosity density in $B$ of $\rho_L = 0.074$ \Lsun
pc$^{-3}$; with a stellar mass density of $\rho_* = 0.042$~\Msun pc$^{-3}$ this
yields a mass-to-light ratio of $(M/L)_B = 0.6$ for the local volume. For the
column, the $B$-band luminosity surface density is $\Sigma_L = 25.6 $ \Lsun
pc$^{-2}$; the stellar mass column density is 35.5 \Msun pc$^{-2}$; hence the
surface mass-to-light in the $B$ band for stellar matter is $(M/L)_B = 1.4$.
As before, the error on the $M/L$ ratio is $\approx 15$\%, and the
mass-to-light ratio is higher by a factor of 1.4 if one includes both stellar
and gaseous disc matter in the surface column density (c.f.\ end of section
2.3).

\subsection{$I$-band disc luminosity and mass-to-light}

The results for the $I$ band are shown in Fig.~\ref{fig:lfMS_I_v2}. As with the
$B$-band, the results are plotted as a function of $M_V$.  For the Heidelberg
sample, colour transformations from $V$ to $I$ were possible for giants only up
to $M_V=0$. Both for the light contributed by MS stars and giants, once more
there is excellent agreement (within 10\%) between the overlapping parts of
each study.

The total volume luminosity density in the $I$ band is $\rho_L = 0.063$ \Lsun
pc$^{-3}$ (this can be computed from the Tuorla sample only).

Again, interesting results emerge when the surface luminosity is computed, as
seen in Fig.~\ref{fig:lfI_tuorla}. We show only the Tuorla data for clarity and
completeness; line symbols are as in Fig.~\ref{fig:lfV_tuorla}
and~\ref{fig:lfB_tuorla}.  Two clear peaks in the surface luminosity
contribution are seen again --- one from giants (mainly clump giants at $M_V
\approx 1$) and the other from turnoff stars ($M_V \approx 4$). Also, brighter
and redder giants, at $M_V \approx -1$ are starting to contribute to the
luminosity. Giants now start to dominate the luminosity budget, contributing
more than half (56\%) of the total light.

\begin{figure}
\begin{center}
\includegraphics[width=0.45\textwidth]{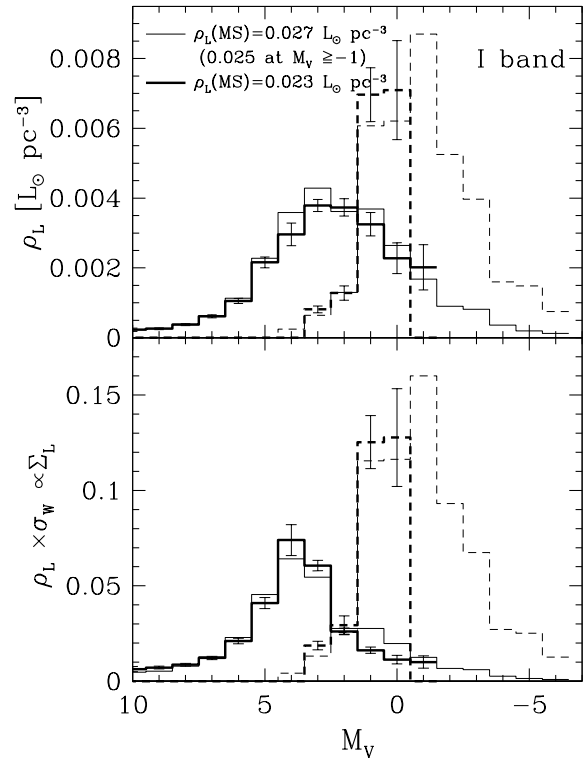}
\end{center}
\caption{Luminosity density in the $I$-band for the local volume (top panel)
and column--scaled (bottom panel). Tuorla results as thin lines, Heidelberg 
results as thick lines with error-bars. Solid lines are for MS stars, dashed
lines for giants. The total $I$-band luminosity generated locally, from the 
Tuorla results, is $\rho_L = 0.063$ \Lsun pc$^{-3}$.}
\label{fig:lfMS_I_v2}
\end{figure}

\begin{figure}
\begin{center}
\includegraphics[width=0.35\textwidth,angle=-90]{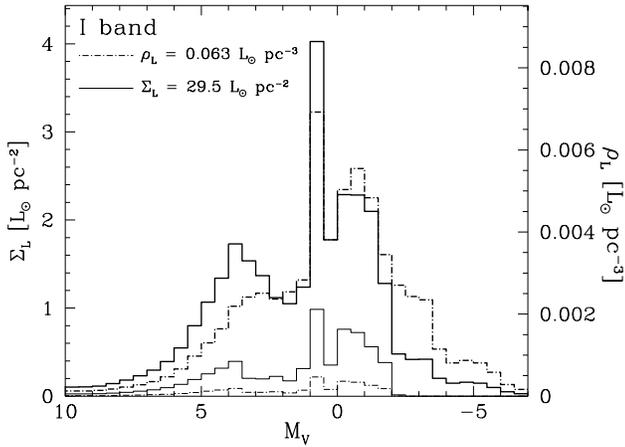}
\end{center}
\caption{Same as Fig.~\protect{\ref{fig:lfV_tuorla}}, but in $I$ band.  Two
clear peaks, associated with turn-off stars and clump giants are again seen.
The contribution of brighter giants, at around $M_V = -2$, is also starting to
have an impact, as one moves to redder bands.  In fact, in the $I$-band in the
solar column, the giants now contribute more than half the light (56\%).}
\label{fig:lfI_tuorla}
\end{figure}

Summarising the $I$-band results and mass-to-light ratios: we obtain a volume
luminosity density in $I$ of $\rho_L = 0.063$ \Lsun pc$^{-3}$; the stellar mass
density is $\rho_* = 0.042$ \Msun pc$^{-3}$ --- this yields an $I$-band
mass-to-light ratio of $(M/L)_I = 0.7$ for the local volume. For the column,
the $I$-band luminosity surface density is $\Sigma_L = 29.5$ \Lsun pc$^{-2}$;
the mass column density is 35.5 \Msun pc$^{-2}$; hence the surface
mass-to-light ratio in the $I$ band is $(M/L)_I = 1.2$ in the column. The error
on the $M/L$ ratio is $\approx 15$\%, and, as before (end of section 2.3) the
mass-to-light ratio is higher by a factor of 1.4 if one includes both stellar
and gaseous disc matter in the local column surface density.

\begin{figure}
\begin{center}
\includegraphics[width=0.45\textwidth]{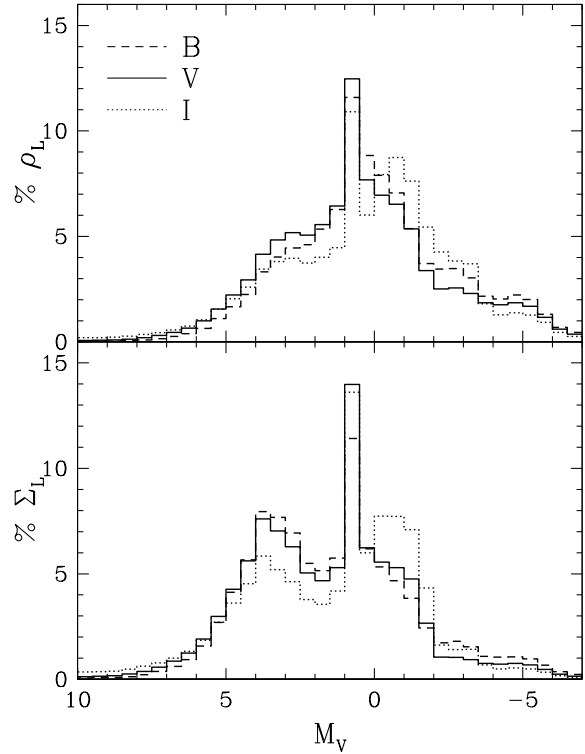}
\end{center}
\caption{Percentage contribution to volume (upper panel) and column (lower
panel) luminosity density in $B, V, I$ band from stars of different
magnitude $M_V$, for the Tuorla disc model.}
\label{fig:lf_BVI}
\end{figure}

As expected, the analysis showed that relative contributions by the two main
factors in the luminosity changes with colour band. In terms of the surface
luminosity $\Sigma_L$, giant stars contribute about 26\% of the light in the
$B$ band, 40\% in $V$ band and 56\% in $I$ band, while only contributing about
0.5\% of the stellar surface mass density. Turnoff stars contribute
significantly to the light in all bands, while upper main sequence and
supergiants ($M_V < -1$) contribute very little to light in all bands.

Finally, in Fig.~\ref{fig:lf_BVI} we compare the luminosity histograms in $B,
V, I$ (in units of percentage contribution to the total volume/column
luminosity) to highlight how the proportion between the to peaks, of main
sequence stars and giants, changes in different bands.

\begin{table}
\caption{Luminosities and mass-to-light ratios for the local volume and local
column, measured in the bands $B$, $V$ and $I$, for the disc's {\it stellar}
material. The error in the luminosity determinations is $\approx 10$\%, while
the error in the mass-to-light ratios is $\approx 15$\%. Note well that the
mass-to-light ratios are for disc matter in stellar form --- the gaseous
components have been explicitly left out. For the local disc, their inclusion
increases the mass-to-light ratios in the local column in all bands by a factor
of 1.4, as discussed at the end of section 2.3; the mass-to-light ratio
increases by a factor of 2.2 in the local volume if the gaseous component is
included.}
\begin{tabular}{lrrrr}
\hline
Property                        & units           &  $B$   &  $V$   & $I$ \\
\hline
volume luminosity, $\rho_L$     &\Lsun pc$^{-3}$ &  0.074 &  0.056 & 0.063 \\
volume mass-to-light ratio      &\Msun/\Lsun      & 0.57   &  0.75  & 0.67  \\
surface luminosity, $\Sigma_L$  &\Lsun pc$^{-2}$ & 25.60  & 24.35  & 29.54 \\
surface mass-to-light ratio     &\Msun/\Lsun      &  1.39  &  1.46  & 1.20 \\
\hline
\end{tabular}
\label{tab:luminosities}
\end{table}

\begin{table}
\caption{Results of the Tuorla study for the volume and column luminosity
and colours of the local thin, think and total disc. The percentage of
luminosity contributed by main sequence stars is also indicated in 
parenthesis, the rest being due mostly to giants.}
\begin{tabular}{lcccc}
\hline
Property & {\it thin disc} & {\it thick disc} & {\it total} \\
\hline
$\rho_B$ [\Lsun pc$^{-2}$]   & 0.072 (82\%) & 0.002 (53\%) & 0.074 (81\%) \\
$\rho_V$ [\Lsun pc$^{-2}$]   & 0.054 (67\%) & 0.002 (41\%) & 0.056 (66\%) \\
$\rho_I$ [\Lsun pc$^{-2}$]   & 0.060 (43\%) & 0.003 (29\%) & 0.063 (43\%) \\
$\Sigma_B$ [\Lsun pc$^{-2}$] & 22.10 (82\%) & 3.50 (53\%) & 25.60 (78\%) \\
$\Sigma_V$ [\Lsun pc$^{-2}$] & 19.92 (64\%) & 4.43 (41\%) & 24.35 (60\%) \\
$\Sigma_I$ [\Lsun pc$^{-2}$] & 22.90 (48\%) & 6.64 (29\%) & 29.54 (43\%) \\
$\mu_B$ [mag as$^{-2}$]      & 23.67 & 25.67 & 23.51 \\
$\mu_V$ [mag as$^{-2}$]      & 23.14 & 24.78 & 22.93 \\
$\mu_I$ [mag as$^{-2}$]      & 22.30 & 23.65 & 22.03 \\
$(B-V)_\rho$                 & 0.32 & 0.89 & 0.34 \\
$(V-I)_\rho$                 & 0.81 & 1.13 & 0.82 \\
$(B-V)_\Sigma$               & 0.53 & 0.89 & 0.58 \\
$(V-I)_\Sigma$               & 0.84 & 1.13 & 0.90 \\
\hline
\end{tabular}
\label{tab:details}
\end{table}

\section{Comparison with theory}

Population synthesis models predict well-defined relations between the colours
and the stellar mass-to-light ratio (M$_*$/L) of composite stellar
populations, with a zero--point depending on the assumed stellar Initial Mass
Function (Bell \& de Jong 2001; Portinari, Sommer--Larsen \& Tantalo 2004).
Fig.~\ref{fig:colourML} compares our results for the solar cylinder to the
theoretical relations by Portinari \etal (2004) computed for different IMFs.
The classic Salpeter IMF (extended down to 0.1~\Msun, see Portinari \etal 2004)
predicts in this comparison a too large M$_*$/L --- but see below --- while the
Kennicutt IMF is too ``light''.  Our (SC) values are instead in very good
agreement with the predictions for the Kroupa (1998) and Chabrier (2001) IMFs,
which were derived from Solar Neighbourhood studies --- a successful
consistency check.

Of course, both the derivation of the stellar IMF in the local field and our
estimate of the surface brightness and colours, ultimately rely on the same
type of data, namely star counts in the Solar Neighbourhood; yet the comparison
is not just tautological, as on one hand we have star counts analyzed and
reproduced with a calibrated model for the vertical structure of the disc, on
the other hand the results of population synthesis models assuming an IMF like
the local one.  And the mass and luminosity contributions are fairly
independent, coming from stars of very different magnitude ranges, as apparent
by comparing Fig.~\ref{fig:massmodel} (mass distribution) to
Fig.~\ref{fig:lf_V} through~\ref{fig:lfI_tuorla} (luminosity distributions),
the mass and luminosity contributions are fairly independent, coming from stars
of very different magnitude ranges.

\begin{figure}
\includegraphics[width=0.50\textwidth]{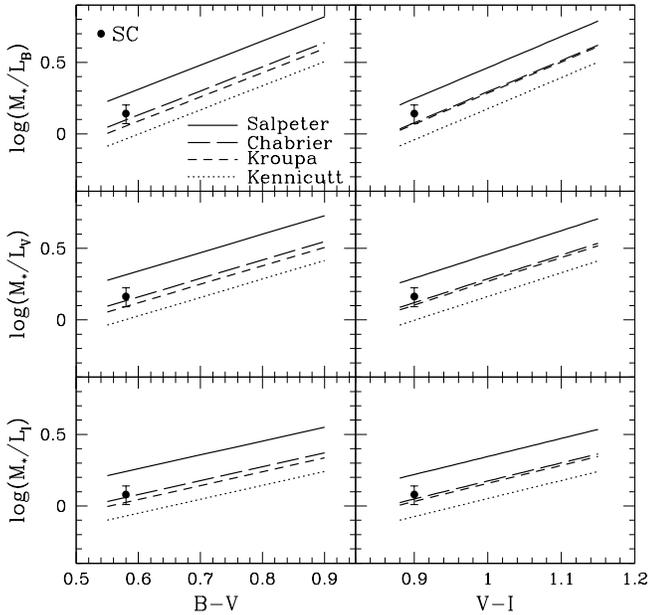}
\caption{Location of the Solar cylinder with respect to the colour--M$_*$/L
relations predicted by stellar populations synthesis models for different
Initial Mass Functions. The results from the Tuorla Galactic model match very
well with the theoretical predictions for the Kroupa and Chabrier IMFs, which
are representative of the Solar Neighbourhood.}
\label{fig:colourML}
\end{figure}

In the previous sections and in Table~\ref{tab:luminosities} the stellar
M$_*$/L was determined on the base of the ``counted'' stellar material, with a
surface density of $\Sigma_* =35.5$ \Msun~pc$^{-2}$ according to the Tuorla
mass model; the total surface density of baryons, including the gas component,
is about 49 \Msun~pc$^{-2}$ (see Fig.~\ref{fig:massmodel}). This is in good
agreement with the dynamical estimates of the surface density of the local disc
by e.g.\ Kuijken \& Gilmore (1991), Flynn \& Fuchs (1994), Bienaym\'e \etal
(2006). It is also compatible, though on the lower end of the uncertainty
range, with the more recent, Hipparcos based dynamical estimate of the surface
density 56$\pm$6~\Msun~pc$^{-2}$ (Holmberg \& Flynn 2000, 2004). This was
derived under the assumption of a spherical dark halo; if the halo is somewhat
flattened the surface density is lower, since dynamically the best determined
quantity is $K_{1.1}$, i.e.\ the total vertical force within 1.1~pc of height
on the Galactic plane, while the disc surface density is a derived quantity
depending on the adopted dark halo model. The halo of the Milky Way is close to
spherical (Ibata \etal 2001; Johnston, Law \& Majewski 2005; Belokurov \etal
2006), but the issue is still debated and an axis ratio of 0.7 may be
compatible with the data (Ibata \etal 2001; Mart{\'\i}nez--Delgado \etal 2004;
Helmi 2004).  So we can consider 56$\pm$6 as an upper limit to the total
surface density, which allows for at most 6--12~\Msun~pc$^{-2}$ not ``seen'' in
the visible baryons. By imputing {\it all} of that to undetected extra stellar
matter --- and not, for instance, to uncertainties/errors in the gas
contribution (which indeed are about 50\%), or to a minor dark matter component
--- the stellar surface density would increase from $\sim$36 to
42--48~\Msun~pc$^{-2}$, and M$_*$/L would increase by 15-30\% over the values
given in Table \ref{tab:luminosities}. We note that 30\% is about the
difference between the M$_*$/L ratios typical of a Kroupa/Chabrier IMF and a
Salpeter IMF (see Fig.~\ref{fig:colourML} and e.g.\ Fig.~4 in Portinari \etal
2004).  The Salpeter scaling can thus be seen as corresponding to the maximum
upper limit to M$_*$/L allowed by the dynamical mass estimates in the Solar
Neighbourhood.

\begin{figure}
\begin{center}
\includegraphics[width=0.45\textwidth]{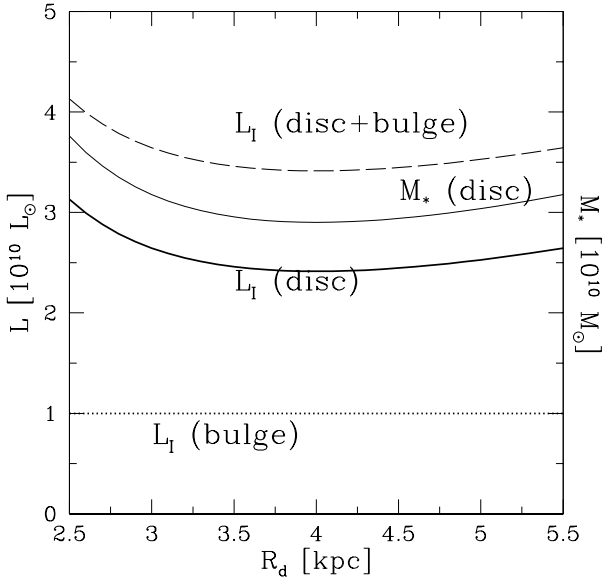}
\end{center}
\caption{Total $I$ band luminosity of the Milky Way inferred as a function of
the assumed disc scale-length, $R_d$. The total luminosity is fairly
insensitive to the adopted scale-length, at least in the range 2.5 to 5 kpc.}
\label{fig:MW-TFa}
\end{figure}

\section{From the Solar cylinder to the Milky Way}

From the surface luminosity (or density) at the Solar radius one can infer the
total luminosity (or mass) of the Galactic disc by assuming an exponential
radial profile for the light (or mass); the result is surprisingly {\it
insensitive} to the assumed scale-length $R_d$, for a plausible range of
scale-lengths (i.e. 2.5--5~kpc; as shown by Sommer--Larsen \& Dolgov, 2001).
The function $L_{disc}(R_d)$ --- or $M_{*,disc}(R_d)$ --- has a very broad
minimum around $R_d \approx R_{\odot}/2$, and for $R_{\odot} = 8$ kpc its shape
is as in Fig.~\ref{fig:MW-TFa}.

One caveat we should consider is that the Solar cylinder probes an inter-arm
region, so we must allow for the spiral arm contrast in deriving the
azimuthally averaged surface brightness at the Solar radius.  We will
henceforth focus on the $I$ band as this is most typical for Tully--Fisher (TF)
studies, and spiral arms are not expected to be very prominent. From the Near
Infrared (NIR) Galactic model of Bissantz \& Gerhard (2002), we estimate that
spiral arms enhance by only 10\% the azimuthally averaged NIR surface
brightness at the solar radius, and by at most 13\% the overall disc luminosity
(see also Gerhard 2002; Drimmel \& Spergel 2001); in the $I$ band the effect of
spiral arms should be of comparable magnitude as the colour contrast is not
significant (Rix \& Zaritsky 1995, and references therein; Grauer \& Riecke
1998).  Fig.~\ref{fig:MW-TFa} thus shows the total $I$ band luminosity and
stellar mass of the Galactic disc inferred from the local surface brightness
and density, including a 10\% correction for spiral arms.  The total disc
luminosity is {\mbox{L$_{I,disc} \sim 2.5-3 \times 10^{10}$~L$_{\odot}$}}.

We must further add the bulge contribution to get the total luminosity of the
Milky Way, to be compared to external spirals.  The bulge luminosity in the NIR
is $\sim 10^{10}$~L$_{\odot}$ (Kent et~al.\ 1991; Gerhard 2002); we assume the
same value in $I$ band, which is probably an overestimate as the bulge is
mostly composed of red old populations so its NIR luminosity should be larger
in proportion than the $I$ band one. The total $I$ band luminosity of the Milky
Way is thus $\sim 3.8 \pm 0.6 \times 10^{10}$~L$_{\odot}$ (where the error is
estimated adding in quadrature an uncertainty of $\pm$0.3 from the disc
scalelength in Fig.~\ref{fig:MW-TFa} and a 10\% uncertainty estimated for our
determination of the local surface luminosity in Section~3), corresponding to
$M_{\rm I} \sim -22.3$. With a circular speed of $\sim 220 \pm 20$~km/sec (the
current IAU standard), the Milky Way turns out to be underluminous with respect
to the TF relation defined by external spirals (Fig.~\ref{fig:MW-TFb}).
Although the discrepancy is not dramatic (about 1 $\sigma$) when considering
the scatter in the TF relation and the errors in the Milky Way values, it may
indicate a problem with the zero--point of the TF relation or with the stellar
mass-to-light ratio of disc galaxies.

\begin{figure}
\begin{center}
\includegraphics[width=0.45\textwidth]{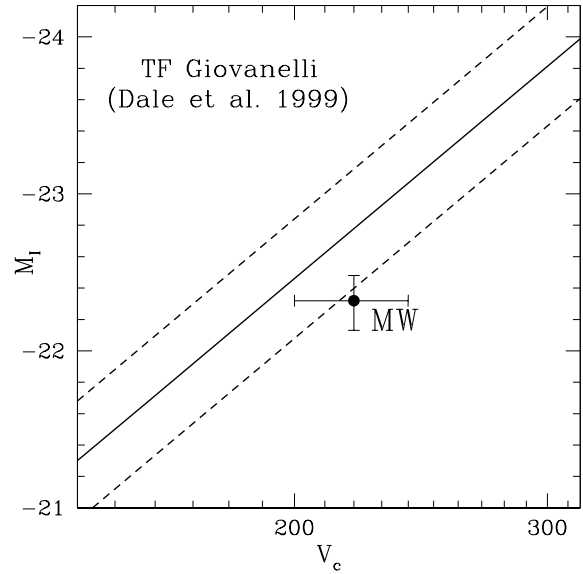}
\end{center}
\caption{Location of the Milky Way with respect to the $I$ band Tully-Fisher
relation (solid line). The dashed lines indicate the 1-$\sigma$ scatter in the
Tully-Fisher relation.}
\label{fig:MW-TFb}
\end{figure}

\subsection{Confirming the offset: stellar/baryonic mass}

In this section we show that the offset of the Milky Way from the TF relation
is confirmed when considering the TF relation in stellar and baryonic mass.

Fig.~\ref{fig:totalmass} shows the stellar mass of the Galactic disc estimated
as a function of disc scalelength from our local stellar surface density
(35.5~\Msun~pc$^{-2}$, plus a 10\% enhacement due to the spiral arms on the
azymuthal average). For the stellar mass profile scalelengths as short as
2--2.6~kpc have been advocated, based on the global NIR emission or studies of
the local stellar distribution (Freudenreich 1998; Vallenari \etal 2000;
Bissantz \& Gerhard 2002), which correspond to $M_*$(disc)=3.6--5.4~$\times
10^{10}$~\Msun. To estimate the bulge mass, we follow Sommer--Larsen \& Dolgov
(2001) by imposing that the total circular velocity of the bulge+disc
combination at 3~kpc does not exceed the observed value of 200~km/sec (Rohlfs
\etal 1986). For $R_d = 2$ kpc, the stellar disc mass enclosed within the
central 3~kpc is 2.4~$\times 10^{10}$~\Msun, which accounts already for the
bulk of the circular velocity with little room for a significant bulge mass. In
general, for $R_d \leq 2.6$ kpc, bulge masses of $ \leq 1.3 \times
10^{10}$~\Msun\ at most are compatible with the dynamical
constraints. Altogether, the resulting total stellar mass of the Milky Way is
4.85--5.5~$\times 10^{10}$~\Msun for any $2 \leq R_d \leq 5.5$.

The total baryonic mass is obtained by adding the gas mass in the atomic and
molecular phases, about $9.5 \pm 3 \times 10^9$~\Msun\ (Dame 1993, after
applying a 40 percent correction to account for the mass contribution of
helium), out of which $3 \pm 1 \times 10^9$~\Msun\ lies within the solar
circle. This yields a total baryonic mass for the Milky Way of around $6.1 \pm
0.5 \times 10^{10}$~\Msun, of which $4.9 \pm 0.4 \times 10^{10}$ lies within
the solar circle. Our ``back of the envelope'' estimate is comparable with the
estimate of $5.5 \times 10^{10}$ (within the solar circle) obtained from full
models of the NIR emission and gas dynamics in the Milky Way (Gerhard 2002);
the two estimates agree to better than 15\%, which we assume to be the error in
the total mass estimates.

In Fig.~\ref{fig:MW-TFmass} we locate the Milky Way in the observational plane
of the stellar mass TF relation; the offset of about 1~$\sigma$ with respect to
external galaxies is confirmed. With respect to the baryonic TF relation
(McGaugh 2005; Fig.~\ref{fig:MW-TFbar}) the Milky Way lies close enough to the
observed relation when stellar masses are estimated on the base of population
synthesis models (dashed lines) but the offset is very significant when the
stellar $M_*/L$ ratio is assigned with the favoured recipe of minimizing the
scatter in the empirical mass discrepancy -- acceleration relation (and
consequently, in the TF relation itself, see McGaugh 2005 for details; solid
lines).

All in all, the offset between the Milky Way and external galaxies is
confirmed.

\begin{figure}
\begin{center}
\includegraphics[width=0.45\textwidth]{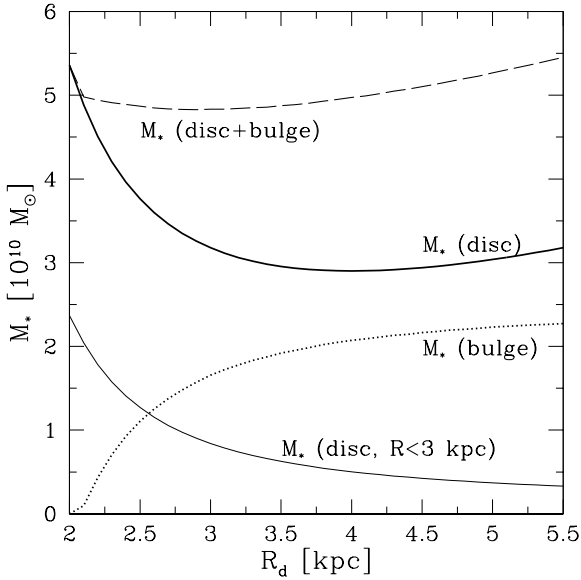}
\end{center}
\caption{Stellar mass in the disc, bulge and total Milky Way inferred 
as a function of the assumed disc scale-length, $R_d$.}
\label{fig:totalmass}
\end{figure}

\begin{figure}
\begin{center}
\includegraphics[width=0.45\textwidth]{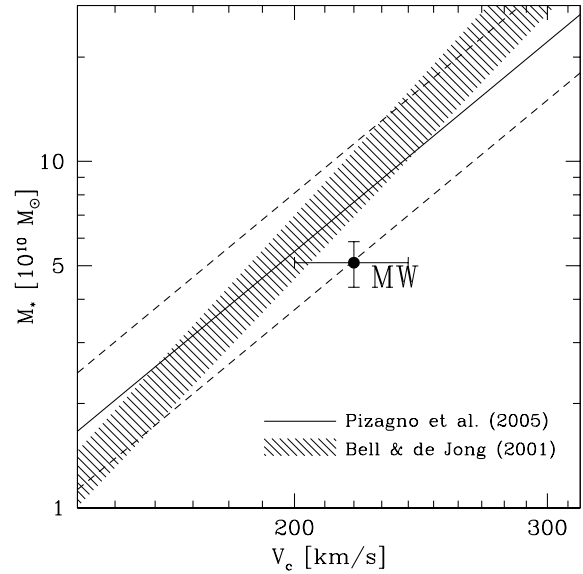}
\end{center}
\caption{Location of the Milky Way with respect to the stellar mass 
Tully-Fisher relation}
\label{fig:MW-TFmass}
\end{figure}

\begin{figure}
\begin{center}
\includegraphics[width=0.45\textwidth]{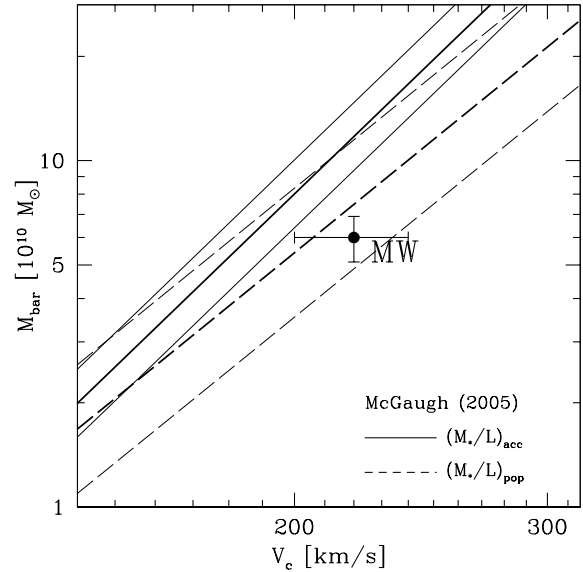}
\end{center}
\caption{Location of the Milky Way with respect to the baryonic Tully-Fisher
relation.}
\label{fig:MW-TFbar}
\end{figure}

\subsection{Discussion}

In the following we discuss possible causes and solutions to this discrepancy.

\begin{itemize}

\item We took our very local value of the surface brightness to infer the
global Milky Way luminosity and mass. The Solar Neighbourhood is an inter--arm
region and, although we did correct for spiral arm enhancement relying on the
current understanding of the spiral structure of the Milky Way, it is possible
that the correction is underestimated. In fact, current models indicate for the
Milky Way significantly weaker spiral arms than is typical for other spiral
galaxies in $K$ band (Rix \& Zaritsky 1995; Drimmel \& Spergel 2001; though
Seigar \& James 1998 also indicate an average arm strength of 10\% );
henceforth we might be presently underestimating the fraction of light in
spiral arms for our own Galaxy.

Conversely, local enhancements of bright stars such as Gould's Belt or the
local arm (Orion Spur), may tend to boost the surface brightness in the Solar
Cylinder --- reinforcing the discrepancy with the TF relation. However the
Heidelberg sample, limited to the nearest 25--50 pc, is free from such
potential ``contaminations'' and the excellent agreement between the Tuorla and
the Heidelberg results provides a consistency check for magnitudes $M_V \geq
-1$ (Fig.~\ref{fig:lfMS_I_v2}); stars brighter than those in the Tuorla study
provide less than 10\% percent of the total $I$ band surface luminosity
(Fig.~\ref{fig:lfI_tuorla}) hence we do not expect Gould's belt or the Orion
spur to induce major offsets. Besides, the fact that our stellar $M_*/L$ is in
excellent agreement with population synthesis predictions for the local stellar
IMF, is an argument against significant local biases.

\item The Milky Way may be a redder and earlier type spiral than the Sbc-Sc
galaxies typically defining the TF relation, requiring a significant colour
correction for the comparison (Kannappan et~al.\ 2002; Portinari et~al.\ 2004);
however, the disc is locally quite blue ($B-V \sim 0.6$, Table
\ref{tab:details}) which argues against major colour and $M_*/L$ offsets
between the Milky Way and Sbc-Sc spirals, which typically have $B-V \sim 0.55$.
Although full chemo--photometric models of the Milky Way are necessary to
explore this possibility, the fact that the offset is found also when
considering the stellar mass and baryonic TF relation, reinforces the
conclusion that it cannot be purely a colour/Hubble type effect.

\item From Fig.~\ref{fig:MW-TFa} it is clear that the inferred total luminosity
of the disc rises sharply if the radial scalelength is shorter than 2.5~kpc. An
$I$ band scalelength as short as 2 kpc would yield a total Milky Way luminosity
of $5.5 \times 10^{10}$~\Lsun\ or {\mbox{$M_I=-22.7$,}} within 0.1~mag 
of the TF locus.
Scalelengths as low as 2--2.2~kpc have been advocated for the NIR emission and
the stellar mass distribution (Section 5.1); however it is likely that in bluer
bands, such as $I$ band, the scalelength is longer due to the colour gradients
typical of galactic discs (de Jong 1996), and in fact longer scalelengths of
3.5--5~kpc are quoted for e.g.\ the $B$ band (de Vaucouleurs \& Pence 1978; van
der Kruit 1986).

More importantly, if its $I$ band scalelength were 2~kpc, the Milky Way would
be a 1.5~$\sigma$ outlier in the scalelength--$V_c$ relation; for its circular
velocity, in fact, external spirals typically have an $I$ band $R_d = 3.5 \pm
1$~kpc (e.g.\ Fig.~8 in Sommer--Larsen, G\"otz \& Portinari 2003). Therefore
the Milky Way, though matching in that case the TF relation, would still be a
non--typical spiral galaxy, from a different point of view.

\item In Fig.~\ref{fig:MW-TFb} we compare the Milky Way to the TF relation by
Dale et~al.\ (1999). Portinari et~al.\ (2004, Appendix A) discuss differences
and offsets among observational TF relations by different groups. Although
lower TF luminosity normalizations do exist in literature, it seems that all
the most recent formulations, based on inverse or bivariate fits to the data,
are in very good agreement. Especially considering that the new Dale
et~al. normalization is 0.1~mag dimmer than the earlier TF by Giovanelli
et~al.\ (1997), there is close agreement with the results of e.g.\ Tully
et~al.\ (1998), Tully \& Pierce (2000), Courteau et~al.\ (2000).  
The Cepheid--distance based TF of
Sakai et~al.\ (2000) is steeper than the others, yet this affects only objects
of smaller size that Milky Way, while for $V_c \sim 200$~km/sec the Sakai TF
luminosities are similar to the Dale \etal and other results. Consequently, the
discrepancy we found is not just a result of the specific TF data set to which
we have compared. Furthermore, the offset is confirmed when considering other
TF planes, like the stellar mass and baryonic TF relations
(Fig.~\ref{fig:MW-TFmass} and Fig.~\ref{fig:MW-TFbar}).

\item 
The TF relations we considered in the previous item correspond to $H_0 \simeq
70$~km~sec$^{-1}$~Mpc$^{-1}$, and/or to the Cepheid HST Key--project distance
scale. A larger value of $H_0$, would be required to make external galaxies
closer and intrinsically fainter, and reducing the Milky Way's offset from the
mean TF. The value of $H_0$ is still debated on the basis of (a) the Cepheid
period--luminosity calibration and its metallicity dependence, (b) the
treatment of extinction, and (c) possible systematic biases and of the
comparison to other distance indicators (Sandage \etal 2006, and references
therein; Ngeow \& Kanbur 2006; Storm \etal (2005); Teerikorpi \& Paturel 2001;
Tully \& Pierce 2000).  However, most of the above mentioned studies point to
longer distance scales and lower values of $H_0 \simeq
60$~km~sec$^{-1}$~Mpc$^{-1}$, which would increase the offset in
Fig.~\ref{fig:MW-TFb}. Dimming the TF relation by 0.5~mag in
Fig.~\ref{fig:MW-TFb} would require an upward revision of $H_0$ by 26\%, or
$H_0 \simeq 90$~km~sec$^{-1}$~Mpc$^{-1}$, which looks beyond the plausible
range presently discussed (but see Tully \& Pierce 2000).

\item Finally, the TF relation is so steep (L $\propto$ V$_c^\alpha$, with
$\alpha$=3--4) that a small error in the circular velocity will result in a
major offset in luminosity. From Fig.~\ref{fig:MW-TFb}, one can see that the
Milky Way would nicely fall onto the TF relation if its circular speed were as
low as $\sim$190~km/sec. Although estimates of the local circular velocity down
to 185~km/sec can be found (e.g.\ Olling \& Merrifield 1998; Dias \&
L$\acute{\rm e}$pine 2005), they are at the lowermost end of the plausible
range (Sackett 1997; Majewski et~al.\ 2006; and references therein) and seem to
be excluded by recent studies of the motion of open clusters (Frinchaboy \&
Majewski 2006).  Besides, direct measurements of the proper motion of
Sagittarius~A (the Galactic centre) favour rotation speeds as high as
235~km/sec (Reid \& Brunthaler 2004) and even values up to 255~km/sec have been
proposed in the literature (Uemura et~al.\ 2000).

Also biases in the observationally determined circular speed of TF spirals are
of concern here, and are related to different tracers or definitions of
circular velocity. The TF relation by Giovanelli et~al.\ (1997) and Dale
et~al.\ (1999) is based on the HI linewidth $W_{50}$, which corresponds very
closely to twice the maximum velocity of the optical disc $V_{max} \sim
V_{2.2}$ (Courteau 1997); and the circular speed at the solar radius discussed
above cannot be but a lower limit to $V_{max}$ for the Milky Way
disc. Henceforth, though we can expect offsets of order 10~km/sec between
$W_{50}$/2 and $V_{max}$ around $V_{max} \sim$200~km/sec (Kannappan et~al.\
2002), the effect does not seem to be large enough to account for the
discrepancy in Fig.~\ref{fig:MW-TFb}. Besides, as mentioned above, different
$I$ band TF relations defined by different groups, with different kinematic
tracers (Dale et al.\ 1999; Tully et~al.\ 1998; Courteau et~al.\ 2000; Sakai
et~al.\ 2000) are in very good agreement on the luminosity level of galaxies
with Milky--Way like rotation velocities.

\item Dust is unlikely to be responsible for the discrepancy, since both the TF
relation and our Galactic disc model are dust corrected.  As our Galactic disc
study is based on quite nearby stars, it is hardly plausible that dust
corrections have been underestimated by as much as 50\% (which is the increase
in luminosity required to bring the Milky Way in perfect agreement with the TF
relation). Likewise, if dust corrections for external spirals have been
systematically overestimated (artificially brightening the zero--point of the
TF relation), it is unrealistic that this is by as much as 0.4~mag.
\end{itemize}

Although many interpretations and possible solutions of the Milky Way's offset
relative the the mean Tully-Fisher relation have been discussed here, our
findings do suggest a possible problem with the luminosity zero--point of the
TF relation, and/or with the stellar mass-to-light ratio of disc galaxies. A
similar conclusion is suggested also by cosmological semi--analytic models or
simulations of galaxy formation (Dutton et al.\ 2006; Gnedin \etal 2006;
Portinari \& Sommer--Larsen 2006).  The issue certainly deserves further
investigation.

\section{Conclusions}

We have presented new estimates of the $B, V, I$ luminosity density, colours
and mass-to-light ratios of the ``Solar volume'' and ``Solar cylinder'' , i.e.\
the local Milky Way disc, based on Hipparcos/Tycho data and the use of a
calibrated model for the vertical structure of the disc to correct from volume
to column quantities (the Tuorla study). Excellent agreement is found with the
parallel Heidelberg study determining the local luminosity density on
completely empirical grounds, based on the complete CNS4 catalogue of nearby
stars, at least over the (more limited) magnitude range probed by this latter
study.

Our determination of the $B, V, I$ colours and stellar mass-to-light ratios for
the Solar cylinder is in excellent agreement with theoretical expectations,
from stellar population synthesis, for a Kroupa/Chabrier IMF representative of
the Solar Neighbourhood. The uncertainty in the dynamical estimate of the local
surface mass density indicate the Salpeter IMF as the extreme upper limit
allowed for the local stellar $M_*/L$; but the Kroupa/Chabrier scaling remains
most plausible.

Surface, or column, luminosities and colours provide important constraints for
chemo--photometric models of the Solar Neighbourhood and the Milky Way, and
allow to compare the Milky Way to external spirals. We have reconstructed from
the local disc brightness the total $I$ band luminosity of the Milky Way, which
appears to be low with respect to the TF relation; although the
discrepancy is not dramatic (about 1~$\sigma$), we discussed the implications
and possible ways out.  On the other hand, were the Milky Way simply a
1~$\sigma$ outlier from the TF relation (nothing to upset the Copernican
principle), it is still worth to underline such offset as our Galaxy is usually
taken as the paradigm for disc galaxies in general.

We are presently refining our $I$--band luminosity estimate with the aid of
DENIS and other star counts, and defining better our errorbars; we are also
extending our surface brightness determination to other photometric bands,
especially in the infrared utilizing DENIS and 2MASS.

\section*{Acknowledgments}

We thank Neill Reid for kindly allowing us to use his $UBVRI$ dataset on nearby
stars and the anonymous referee for helpful comments. We are very grateful to
the Academy of Finland for considerable financial support (grants nr.~206055
and~208792). LP further acknoweledges the support of a EU Marie Curie
Intra-European Fellowship under contract MEIF-CT-2005-010884.  CF thanks Mount
Stromlo Observatory for its kind hospitality, where part of this research was
carried out.

\end{document}